\documentclass[runningheads]{llncs}
\usepackage{graphicx}
\usepackage{hyperref}

\usepackage{xcolor}
\usepackage{subcaption}
\usepackage{paralist}
\usepackage[misc,geometry]{ifsym} 

\begin{document}
\title{CounQER: A System for Discovering and Linking Count Information in Knowledge Bases}
%
\titlerunning{CounQER: A System for Discovering and Linking Count Information in KBs}
%
\author{Shrestha Ghosh\inst{1}$^\textrm{\Letter}$
\and
Simon Razniewski\inst{1}
\and
Gerhard Weikum\inst{1}
}
\authorrunning{S. Ghosh et al.}
%
\institute{Max Planck Institute for Informatics, Saarland Informatics Campus, 66123 Saarbruecken, Germany\\
\email{\{ghoshs, srazniew, weikum\}@mpi-inf.mpg.de}}
\maketitle              
\begin{abstract}
Predicate constraints of general-purpose knowledge bases (KBs) like Wikidata, DBpedia and Freebase are often limited to subproperty, domain and range constraints. In this demo we showcase CounQER, a system that illustrates the alignment of \textit{counting predicates}, like \texttt{staffSize}, and \textit{enumerating predicates}, like \texttt{workInstitution$^{-1}$}. In the demonstration session, attendees can inspect these alignments, and will learn about the importance of these alignments for KB question answering and curation. CounQER is available at \url{https://counqer.mpi-inf.mpg.de/spo}.  
\keywords{Knowledge bases  \and semantics \and count information.}
\end{abstract}
\section{Introduction}
\textbf{{Motivation and problem.}} 
Detecting inter-predicate relations in Knowledge Bases (KBs) beyond inheritance can lead to a better semantic understanding that can be leveraged for important tasks such as KB curation and question answering (QA). In this work we focus on set predicates and their alignment. Set predicates describe the relation between an entity and a set of entities through two variants - i) \textit{counting predicates} which relate an entity to a count (of a set of other entities) and, ii) \textit{enumerating predicates} which relate an entity to multiple entities.  

Consider a list of counting predicates, \{\texttt{numberOfChildren}, \texttt{staffSize}\}, which take only integer count values as objects and a list of enumerating predicates, \{\texttt{child}, \texttt{employer$^{-1}$}, \texttt{workInstitution$^{-1}$}\}, which take only entity values. 
Identifying set predicates pairs across the two variants, aligned by their semantic relatedness, such as \{\texttt{numberOfChildren} $\leftrightarrow$ \texttt{child}\}, \{\texttt{staffSize} $\leftrightarrow$ \texttt{employer$^{-1}$}\}, \{\texttt{staffSize} $\leftrightarrow$ \texttt{workInstitution$^{-1}$}\}, has two major benefits. 
\begin{enumerate}
    \item\textit{KB curation} - We can discover incompleteness and/or inconsistencies in KBs through alignments~\cite{razniewski2016but,zaveri2016quality}. For instance, if the value of \texttt{numberOfChildren} exceeds the count of \texttt{child} entities of a subject, then the \texttt{child} statements for that subject may be incomplete. Alternately, if the value of \texttt{numberOfChildren} is less than the count of \texttt{child} entities, there may be inconsistent enumerations. An empty instantiation is also an indication of incompleteness.
    \item\textit{QA enhancement} - Set predicate alignments can aid in KB query result debugging and enrichment~\cite{bast2015more,diefenbach2019qanswer}. Even in an event of empty result, for instance, when an entity has no \texttt{numberOfChildren} predicate instantiations, but, has \texttt{child} predicate instances, we can enumerate the object entities of \texttt{child} instead.
    
    Set predicate alignments 
    highlight the variation in predicate usage for the same concept. For instance, the \texttt{staffSize} of an entity has related results on employees through \texttt{employer$^{-1}$} as well as \texttt{workInstitution$^{-1}$}. 
\end{enumerate}

\noindent\textbf{{Approach.}} CounQER (short for ``\underline{\textbf{Coun}}ting \underline{\textbf{Q}}uantifiers and \underline{\textbf{E}}ntity-valued P\underline{\textbf{R}}\-edicates'') uses a two-step approach. First it identifies the counting and enumerating predicates with supervised classification and then aligns set predicate pairs, one from each variant, according to ranking methods and statistical and lexical metrics. For further details refer to~\cite{ghosh2020uncovering}. 
The classification and alignment steps are executed offline. We use the obtained results in our demonstrator for count-related SPO queries on three KBs\footnote{{\url{https://tinyurl.com/wikidata-truthy}, \url{https://tinyurl.com/dbpedia-mappings}, \url{https://tinyurl.com/dbpedia-raw}}} - Wikidata-truthy and two variants of DBpedia based on mapped and raw extractions.

\begin{figure}[t]
    \centering
    \includegraphics[width=\textwidth]{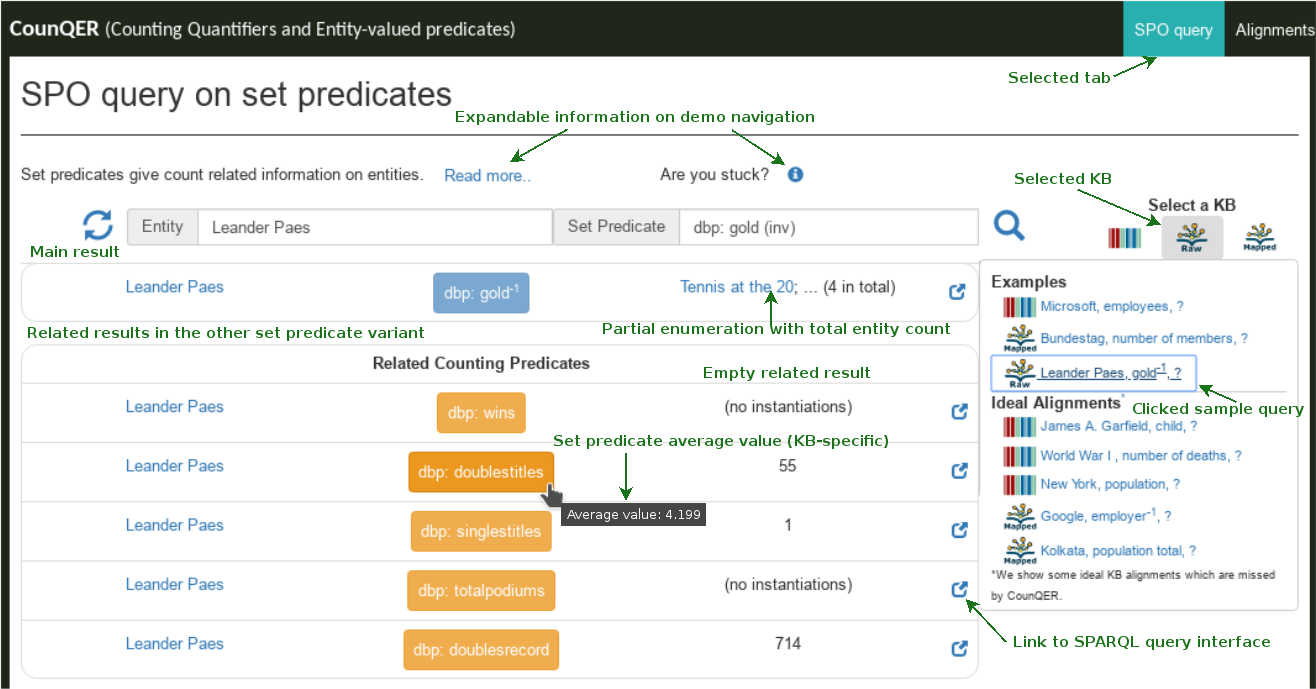}
    \caption{The interface for SPO queries showing results on an example query.}
    \label{fig:interface_spo}
\end{figure}
Fig.~\ref{fig:interface_spo} shows the interface with results on an example query on the DBpedia-raw KB. The query is on the events where the entity, \textit{Leander Paes}, wins gold (\texttt{dbp: gold$^{-1}$}). The main result (set predicate in blue) is succeeded by related results on ranked and aligned set predicates (in orange). Enumerations expand on hovering and we show up to 1000 enumerations. A user can check the actual query fired for each row by following the link to the SPARQL query endpoint. Also, KB-specific predicate statistics show up while hovering over the predicate buttons. On clicking a set predicate from the related results, a new query is fired on the same subject and the new clicked predicate.

\begin{figure}[t]
    \centering
    \includegraphics[width=\textwidth]{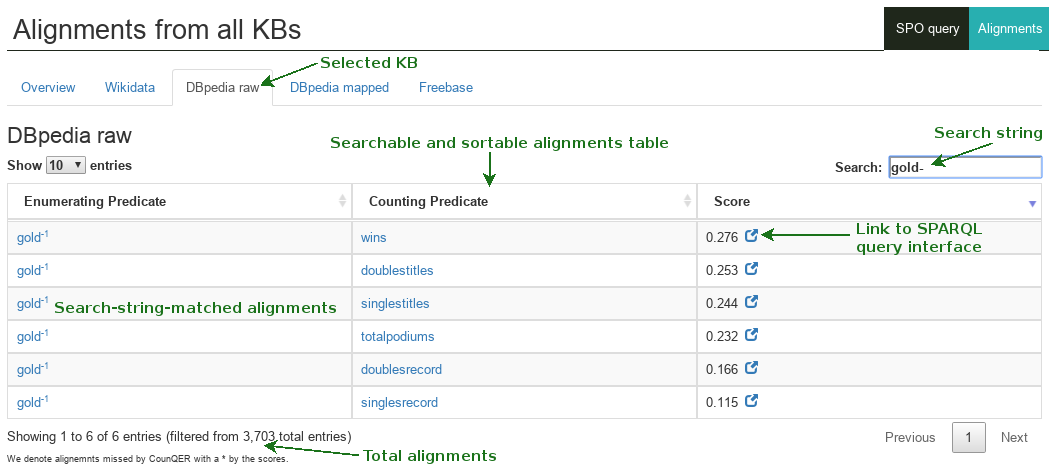}
    \caption{Interface for viewing KB-specific alignments.}
    \label{fig:interface_top_align}
\end{figure}
The complete ranked list of set predicate alignments for the three KBs as well as Freebase can be viewed as in Fig.~\ref{fig:interface_top_align}. Here too, we provide links to the SPARQL endpoint showing the subjects that have populated facts for the alignments.
\medskip

\noindent\textbf{{Related Work.}} Schema and ontology alignment is a classic problem in data integration, which in the semantic web is tackled by focusing on the dynamics of entity relations across ontologies
~\cite{shvaiko2013ontology}. 
Subset and equivalence relation alignment is one of the popular approaches to ontology alignment~\cite{koutraki2017online}. The problem of aligning enumerations with counts, which we address is atypical since most approaches do not target completeness and correctness of KBs~\cite{paulheim2017knowledge}.
Even though between $5\%$ to $10\%$ of questions in popular TREC QA datasets deal with counts~\cite{mirza2018enriching}, QA systems like AQQU~\cite{bast2015more} and QAnswer~\cite{diefenbach2019qanswer} only perform ad-hoc count aggregation function to deal with typical count questions, which start with \textit{``How many..?''}. 

\section{System Description}

\textbf{{SPO query.}}
The SPO query function provides two input fields, \textit{Entity} and \textit{Set Predicate}, and a KB selection button. The first field provides real-time entity suggestions from the selected KB, based on the input prefix, to the user to choose from. 
Next, the user selects a set predicate from the set predicate input field. The predicate choices are KB-specific and ordered by i) whether they are populated and have alignments, ii) they are populated but without alignments, and iii) they are unpopulated. 

Upon execution, the input parameters are sent to our server, where we determine the variant of the user-selected set predicate - counting or enumerating. Then from the KB-specific alignments containing the queried predicate, we shortlist the top-five highest scoring pairs to obtain related set predicate facts. If there are no alignments we do not generate any related query. The server then fires the main query to the SPARQL endpoint of the corresponding KB followed by the SPARQL queries for the aligned set predicates, if present. Once these results are obtained, the server returns the results along with KB-specific predicate statistics, \emph{i.e.,} the average value that the counting predicates take and the average number of entities per subject that the enumerating predicates take.


\noindent\textbf{{Alignments.}}
CounQER provides an option of viewing all alignments across the four KBs along with their alignment scores. A user can go through the list ordered by the alignment score or, use the search bar to filter matching set predicates and view their corresponding alignments. Each alignment has a link to SPARQL query API where the user can view the list of subjects for which the predicate pair co-occur. 

\noindent The main features of the interface are as follows.
\begin{compactenum}
    \item\textit{Predicate suggestions} - Set predicates are ordered based on whether they are populated for the selected entity and whether alignments exist for them.
    \item\textit{Empty results} - If the main query returns an empty result, but, the predicate has populated alignments, CounQER shows the related results. Conversely, if the set predicate in the main query is populated and alignments exist for this predicate, we show the related results regardless of them being empty, thus highlighting potential incompleteness in the KB, w.r.t the queried entity.
    \item\textit{Links to SPARQL queries} - Every row in the results contains a link to the SPARQL endpoint, which a user can follow to check the actual query that was fired and also view enumerations of size more than 1000. Alignment tables also link to the SPARQL endpoint with queries which list subjects for which the set predicate pair co-occur.
\end{compactenum}

We also show some manually added ideal alignments, \textit{i.e.,} the alignments which are present in the investigated KBs but missed by the automated CounQER methodology. These alignments are also present in the table with a fictitious score between [0.9-1]. 
\begin{figure}[t]
    \centering
    \includegraphics[width=0.6\textwidth]{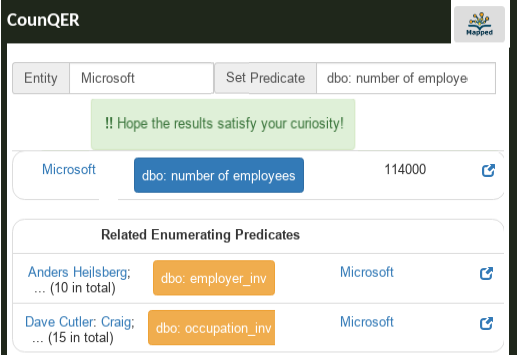}
    \caption{Query results on the number of employees in Microsoft (DBpedia-mapped).}
    \label{fig:user_exp_1}
\end{figure}

\begin{figure}[t]
    \centering
    \includegraphics[width=0.6\textwidth]{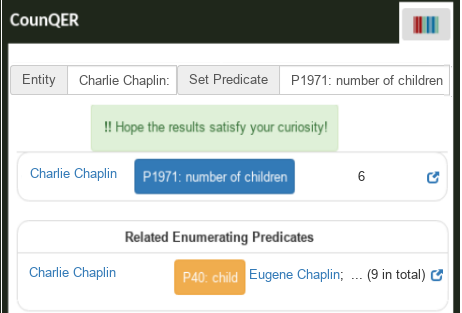}
    \caption{Query results on the number of children of Charlie Chaplin (Wikidata).}
    \label{fig:user_exp_2}
\end{figure}

\section{Demonstration Experience}
\textbf{{Scenario 1 - QA.}} In a query about the number of employees at Microsoft, CounQER finds the main result from the queried KB, DBpedia-mapped, to be $114,000$ employees. In addition, CounQER returns instantiated facts on interesting enumerating predicates, such as, \texttt{employer$^{-1}$} and \texttt{occupation$^{-1}$} (see Fig.~\ref{fig:user_exp_1}).

\begin{figure}[t]
    \centering
    \includegraphics[width=0.6\textwidth]{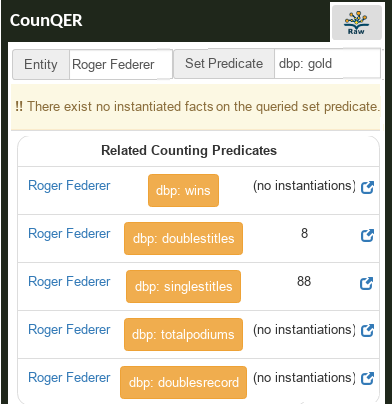}
    \caption{Query results on golds won by Roger Federer (DBpedia-raw).}
    \label{fig:dbpedia_raw_eg}
\end{figure}
\smallskip
\noindent\textbf{{Scenario 2 - KB curation.}} Consider the example in Fig.~\ref{fig:user_exp_2}, where the user searches for the number of children of the British comic actor, Charlie Chaplin. The alignment results reveal inconsistent information in Wikidata-truthy. While the value for \texttt{number of children} is $6$, there are 9 statements for the enumerating predicate \texttt{child}. 

Next, we investigate the winning titles of Roger Federer in 
DBpedia-raw (Fig.~\ref{fig:dbpedia_raw_eg}). Even though a query on the golds won by Federer returns no main results, unlike the query on the golds won by Leader Paes in Fig.~\ref{fig:interface_spo}, the counting predicates \texttt{doublestitles} ($2^{\textit{nd}}$) and \texttt{singlestitles} ($3^{\textit{rd}}$) give the number of doubles and singles titles won by Federer. 

\section{Conclusion}
We demonstrate how set predicate alignments highlight redundancies in the KB schema, enhance question answering by providing supporting counts and/or enumerations and help in KB curation. Utilizing KB alignments to automatically flag inconsistent SPO facts for resolution and highlight SPO facts needing completions is a possible future work. Analysing multi-hop alignments and extending KB alignments towards open information extraction is also worth exploring.

\bibliographystyle{splncs04}
\bibliography{references}
\end{document}